\documentclass[useAMS,usenatbib,referee,epsfig,epsf]{mn2e}
\usepackage{times,epsfig}

\title[First Kepler results on compact pulsators III]
{First \emph{Kepler} results on compact pulsators III: Subdwarf B stars with
V1093~Her and hybrid (DW~Lyn) type pulsations.}

\author[M.D. Reed et al.]{
 M.~D.~Reed,$^1$\thanks{E-mail:MikeReed@missouristate.edu}
S.~D.~Kawaler,$^2$ R.~H.~\O stensen,$^3$ S.~Bloemen,$^3$, A.~Baran,$^{2,4}$
J.~H.~Telting,$^5$ \newauthor R.~Silvotti,$^6$ S.~Charpinet,$^7$  
A.~C.~Quint,$^1$ G.~Handler,$^8$
R.~L.~Gilliland,$^{9}$ W.~J.~Borucki,$^{10}$\newauthor D.~G.~Koch,$^{10}$ 
H.~Kjeldsen,$^{11}$ and J.~Christensen-Dalsgaard$^{11}$\\
 $^1$Department of Physics, Astronomy and Materials Science,
 Missouri State University, 901 S. National, Springfield, MO 65897, USA \\
 $^2$Department of Physics and Astronomy, Iowa State University, Ames, IA 50011, USA\\
$^3$Instituut voor Sterrenkunde, Katholieke Universiteit Leuven, Celestijnenlaan 200 D, 3001 Leuven, Belgium\\
$^4$Krakow Pedagogical University, ul. Podchor\c{a}\.{z}ych 2,30-084 Krak\'{o}w, Poland\\
$^5$Nordic Optical Telescope, 38700 Santa Cruz de La Palma, Spain\\
$^6$INAF-Osservatorio Astronomico di Torino, Strada dell'Osservatorio 20, 10025 Pino Torinese, Italy\\
$^7$Laboratoire d'Astrophysique de Toulouse-Tarbes, Universit\'{e} de Toulouse, CNRS, 14 Av. E. Belin, 31400 Toulouse, France\\
$^8$Institut f\"{u}r Astronomie, Universit\"{a}t Wien, T\"{u}rkenschanzstrasse 17, 1180 Wien, Austria\\
$^9$ Space Telescope Science Institute, 3700 San Martin Drive, Baltimore, MD 21218, USA\\
$^{10}$NASA Ames Research Center, MS 244-30, Moffett Field, CA 94035, USA\\
$^{11}$Department of Physics, and Astronomy, Building 1520, Aarhus University, 8000 Aarhus C, Denmark}

\date{Accepted
      Received }

\begin{document}

\maketitle

\begin{abstract}
We present the discovery of nonradial pulsations in five hot subdwarf B
(sdB) stars
based on 27 days of nearly continuous time-series photometry using
the \emph{Kepler spacecraft}.  We find that \emph{every} sdB star cooler than 
$\approx 27\,500\,$K that Kepler has observed (seven so far) is a long-period 
pulsator of the V1093~Her  (PG~1716) class or a hybrid star with both
short and long periods.
The apparently non-binary long-period and hybrid pulsators are
described here.
 The V1093~Her periods range from one to 4.5~h and
are associated with $g-$mode pulsations. Three stars 
also exhibit short periods
indicative of $p-$modes with periods of 2 to 5~m and in addition, these
stars exhibit periodicities between both classes from 15 to 45~m.
We detect the coolest and longest-period V1093~Her-type
pulsator to date, KIC010670103 ($T_{\rm eff}\approx 20\,900\,$K, $P_{\rm max}\approx 
4.5$~h) as well as a suspected hybrid pulsator, KIC002697388 which is extremely cool
($T_{\rm eff}\approx 23\,900\,$K) and for 
the first
time hybrid pulsators which have larger  $g-$mode amplitudes than
$p-$mode ones. All of these pulsators are quite rich with
many frequencies and we are able to apply asymptotic relationships to associate periodicities
with modes for KIC010670103.
\emph{Kepler} data are particularly well-suited for these
studies as they are long-duration, extremely high duty cycle observations
with well-behaved noise properties.
\end{abstract}

\begin{keywords}

Stars: oscillations -- stars: variables --
Stars: subdwarfs

\end{keywords}

\section{Introduction}

Subdwarf B (sdB) stars are horizontal-branch stars with masses
near $0.5 $M$_\odot$, thin ($< 10^{-2} $M$_\odot$) hydrogen shells,
and temperatures from $22\,000$ to $40\,000$~K
\citep{heber,saf94}.
Asteroseismology of pulsating sdB stars can potentially probe the
interior structure and provide estimates of total mass, shell
mass, luminosity, helium fusion cross sections, and coefficients
for radiative levitation and gravitational diffusion. To apply the
tools of asteroseismology, however, it is necessary to resolve the
pulsation frequencies.  This usually requires extensive
photometric campaigns, preferably at several sites spaced in
longitude to reduce diurnal aliasing. 
\emph{Kepler's} space-based vantage point provides
data that are well suited for asteroseismic studies as they are practically continuous,
evenly sampled, and uniform.

Pulsating sdB stars have two classes which seem to form a continuous
sequence in temperature. The short-period
pulsators were discovered in 1997 \citep{kill97} and were
named EC~14026 stars after that prototype. Their official designation is
V361~Hya stars, 
but they are commonly referred to as
sdBV stars. Their periods are typically two to three minutes, but can be as
long as 5 minutes with amplitudes typically near 1\% of their mean brightness
\citep[see][for a review of 23 pulsators]{reed07b}. 
These are $p-$mode pulsators with driving likely produced
by an iron-enhanced $\kappa$-mechanism \citep{charp96,char01}. Recent
work suggests that about 10\% of hot sdB stars belong
to this class \citep{roysurv,royP1}.
Long-period pulsators were discovered
in 2003 \citep{grn03} and referred to as PG~1716 stars, after
that prototype and officially designated as V1093~Her
stars. 
Their periods range from 45~m to 2~h with amplitudes typically
$<$0.1\% though amplitudes up to 0.5\%  of their mean brightness 
have been observed.
These are $g-$mode pulsators with driving
also caused by the iron $\kappa$-mechanism \citep{font} and they are
cooler then the V361~Hya stars, though there may be some overlap.
Hybrid sdBV stars were first discovered in 2006 with the prototype being
DW~Lyn \citep{schuh06}. They all have temperatures
at the V361~Hya -- V1093~Her border, and prior
to these \emph{Kepler} data, 4 were known \citep{schuh06,andy,lutz09,andy2}.
An interesting property
of the hybrids is that they have some uncharacteristically
high-amplitude frequencies for both $p-$ and $g-$mode pulsations.
As an example, in Balloon~090100001 (hereafter BA09) the highest
amplitude $p-$mode is at $2807.5\mu$Hz with a V-band amplitude of 53.34~mma
and the highest amplitude $g-$mode is at $325.6\mu$Hz with an ampitude
of 2.19~mma \citep{andy2}. 
Both of these amplitudes are a factor of ten higher then average
non-hybrid stars.
For the hybrids, the highest-amplitude $p-$ and $g-$mode
pairs of amplitudes are $21.7,3.7$; $53.34,2.19$ (V-band);
$6.6,1.5$; and $18.8,2.2$~mma (milli-modulation amplitudes) 
for HS~0702+6043, BA09, HS2201+2610 and
RATJ0455+1305, respectively \citep{schuh06,andy,lutz09,andy2}. 
BA09, which is the best studied of these stars, also shows
a series of low-amplitude modes
that nearly span the gap between the $g-$ and $p-$mode ranges \citep{andy09}.
A broad review of sdB stars can be found in \citet{heber09}.

In this paper we examine the apparently non-binary $g-$mode sdB pulsators
discovered during the \emph{Kepler Mission} data released in December 2009. 
These have \emph{Kepler Input
Catalog} (KIC) numbers of 007664467 and 010670103 for
the V1093~Her-type and 
 002697388 \citep[identified in the Sloan SEGUE extension survey as 
SDSS~J190907.14+375614.2,][]{yanny}, 003527751 \citep[also in the Sloan SEGUE
survey as SDSS~J190337.02+383612.5,][]{yanny} and 
005807616 \citep[previously identified as
KPD~1943+4058, ][and hereafter KPD~1943]{dow} for the hybrid
stars. Their \emph{Kepler} system magnitudes are listed as
16.45, 16.53, 15.02 15.39, and 14.86, respectively. Spectroscopic
properties of these stars are published in \citet[][hereafter, Paper I]{royP1}
using metal line blanketed LTE models of solar metalicity. They are 
$T_{\rm eff}\,\log g$ of $\approx 26\,800\,$K, $5.17$ for KIC007664467;
$\approx 20\,900\,$K, $5.11$ for KIC010670103; $\approx 23\,900\,$K, $5.32$
for KIC002697388; $\approx 27\,600\,$K, $5.28$ for KIC003527751;
and $\approx 27\,100\,$K, $5.51$ for KPD~1943. KPD~1943 was already suspected to
be a long-period pulsator from pre-\emph{Kepler} observations \citep{silv09}
and we confirm those observations.
The binary $g-$mode sdB pulsators appear in \citet[][ Paper V]{lpsdBVbinP1}.

The \emph{Kepler Mission} science
goals, mission design, and overall performance are reviewed
by \citet{bor10} and \citet{koch10}. Asteroseismology for \emph{Kepler}
is being conducted through the \emph{Kepler Asteroseismic Science Consortium} \citep[KASC;][]{gila}.
These
data were obtained as part of the survey mode, where short-cadence
targets are rotated every month during the first year \citep{gila,gilb}.
The striking advantage to \emph{Kepler} data are the many improvements
over ground-based data, particularly for $g-$mode-regime variability.
From the ground, one can only observe a few pulsation cycles during
night time hours and if multisite observations are obtained, then
differing instrument sensitivities become an issue, and because of weather,
gaps will inevitably appear in the data. Additionally, ground-based
observations have to contend with atmospheric issues, such as transparency
that can often change on the timescales near that of the intrinsic
variability for $g-$mode sdBV stars. As
such, \emph{Kepler} data are really optimal for these types of pulsators
in that nearly gap-free data were obtained at a roughly constant cadence
and with no atmospheric issues. This advantage allows us to detect
many more pulsation frequencies than Earth-bound telescopes can obtain,
even with significant multisite effort.

Other \emph{Kepler} papers in this sequence 
on sdBV stars include Paper I \citep{royP1}, which summarizes
target selection and spectroscopic properties; Paper II \citep{sdBVP1},
which examines the first V361~Hya variable; Paper IV \citep{groot}
which provides a model fit to KPD~1943; and Paper V \citep{lpsdBVbinP1},
which discusses V1093~Her stars in close binaries.

\section{Observations}
This paper describes data obtained by the \emph{Kepler Mission} during
2009. 
The data released to the KASC
compact stars working group (WG11) 
are short cadence data with an average integration time of 58.8~s.
Data on all five pulsators described in this paper 
were obtained during Q2.3 (the third month of the second quarter)
over a 27~d span between BJD~2455064
and 2455091 (20 August - 16 September, 2009) and released to the KASC
on 31 December, 2009.
The temporal frequency
resolution is $0.43\mu$Hz for these data and
during these $\approx 27$ days of observations, a total of 45,210 science
images were scheduled. However, because of  a 22.5~h safing event
and other small glitches, including loss of fine
guidance, $\approx 3\,700$ images were either not
obtained or were contaminated in some way. These images were not used
in our analysis. It has also been
found that the long cadence reading of the CCDs affects the short
cadence data by creating artefact peaks at $n\cdot 566.391\,\mu$Hz. 
For most of these data,
variation frequencies are short of $566.391\,\mu$Hz and for the short periodicities
we were careful to avoid these artefacts. Another
\emph{Kepler} processing step is a contamination-correction, 
which describes what fraction of the flux is from nearby stars in the
4~arcsecond aperture. These contamination fractions are 
87.9, 45.0, 14.9, 8.1, and 33.2\% for
KIC007664467, KIC010670103, KIC002697388, KIC003527751, 
and KPD~1943 respectively.
Some of these fractions are so large (most of the flux for KIC007664467 
is attributed
to nearby stars), that the actual amplitudes (measured as deviations
from the mean flux) may be significantly different, though relative
amplitudes within each star would not be affected.  For this paper, we used uncorrected
flux as suggested in Paper I 
because we found the photometric temperatures in the \emph{Kepler} 
Input Catalogue to be
cooler than those from spectroscopy. Until
the contamination process is sorted out, it is safer to use the raw
fluxes, and a correction can be applied later to determine the intrinsic
amplitudes.

\begin{figure}
\psfig{figure=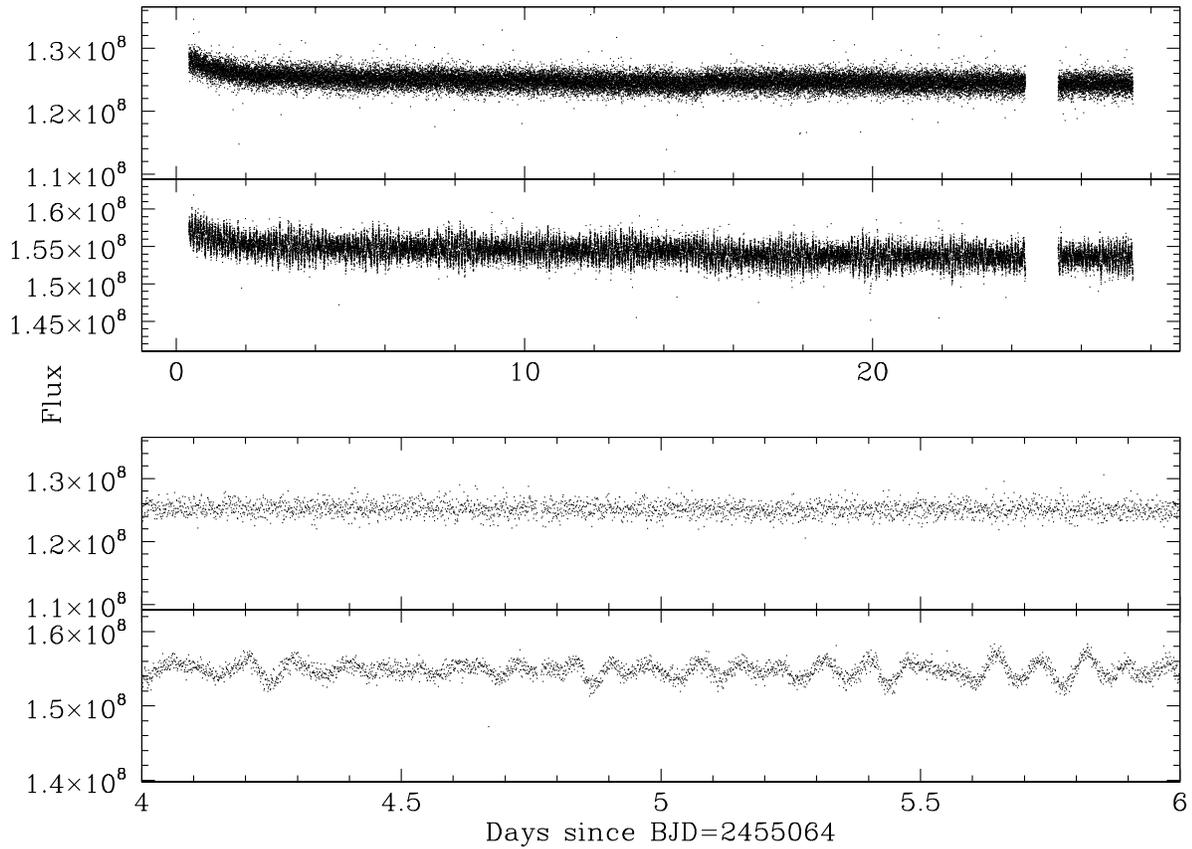,angle=-90,width=\textwidth}
\caption{Light curves of the V1093~Her variables KIC007664467 (top) and 
KIC010670103 (bottom).  
The horizontal axis is in days, and the vertical axis shows 
the raw flux (in counts per second).  
The upper panels show the complete light curves, while the lower ones show 
shorter segments.} \label{fig01}
\end{figure}

\begin{figure}
\psfig{figure=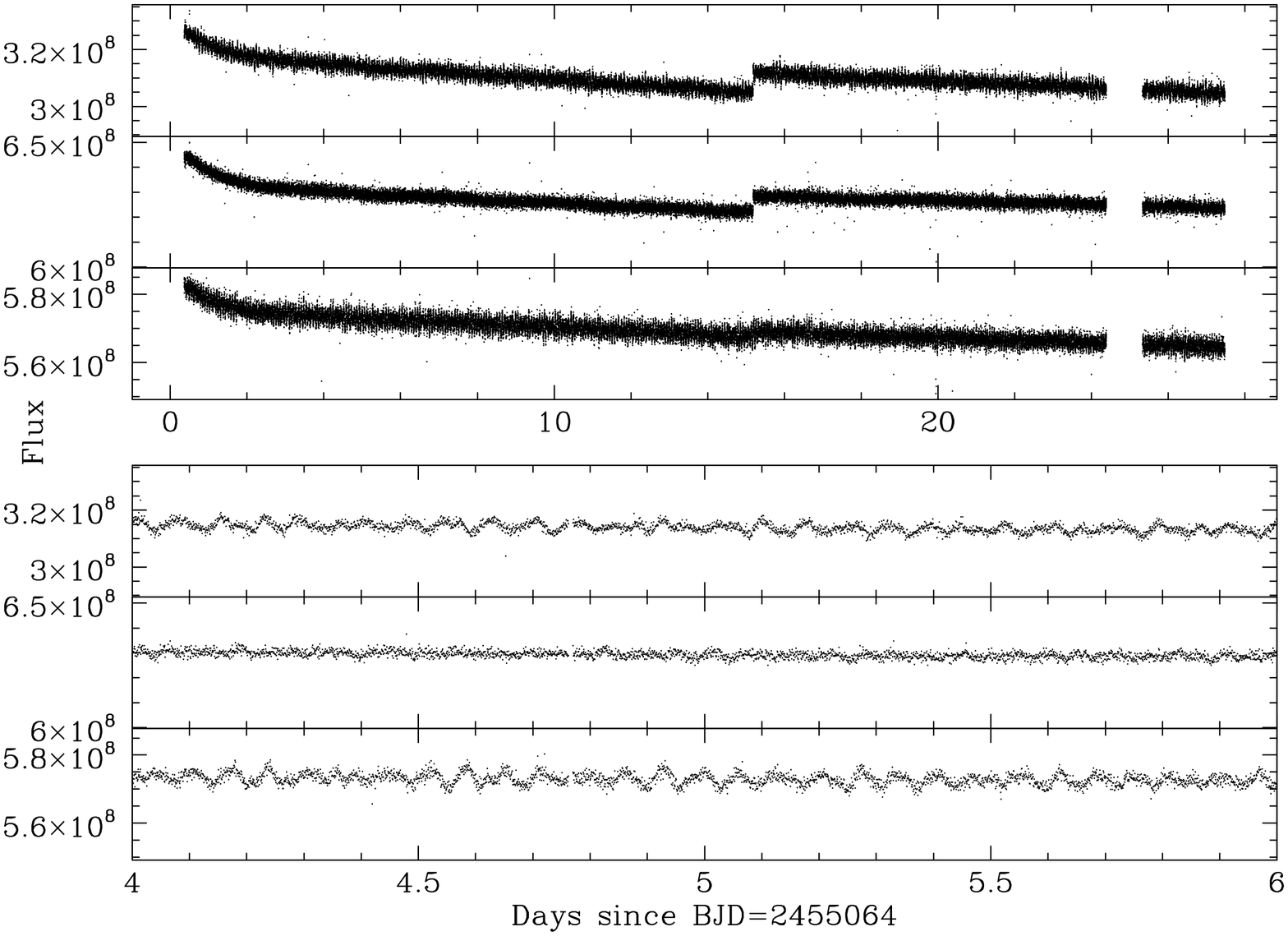,angle=-90,width=\textwidth}
\caption{Same as Fig.\ref{fig01} for the hybrid variables KIC002697388 (top),
KIC003527751 (middle), and KPD~1943 (bottom).  
\label{fig02} }
\end{figure}

\emph{Kepler} data on these stars were provided with times in barycentric 
corrected Julian days, and fluxes which
are shown in Figs.~\ref{fig01} and \ref{fig02}. 
Note the gap during the data that corresponds to a
safing event during day 22. We normalized the flux and
  amplitudes are
given as parts-per-thousand (or milli-modulation amplitude, mma), with 10~mma
corresponding to 1.0\%.

Note that at the time of this writing the pipeline for
reducing \emph{Kepler} short cadence data is still being fine-tuned
and tested. As a result, it is possible that some frequencies and
amplitudes could be affected by the current \emph{Kepler} reduction
pipeline with differing results in subsequent data releases as the
process is improved.

\section{Analyses}
Our analysis proceeded in a straightforward way; we calculated a
temporal spectrum (also known as a 
Fourier transform; FT) which was used for initial estimates of
frequencies and amplitudes. We simultaneously fitted and
prewhitened the data using
non-linear least-squares (NLLS) software beginning with the
highest amplitude frequency until we reached the
$4\,\sigma$ detection limit or unresolved frequencies, which we
were unable to fit. 

We determined $4\sigma$  detection limits
using apparently variation-free frequency regions surrounding
those with periodicities.
For KIC007664467 we used regions bracketing the variations at 
40-80, and 300-500~$\mu$Hz which gave a detection limit of $0.31\,$mma.
For KIC010670103 the bracketing regions used were
20-50 and 205-235, which gave a limit of 0.24~mma.
 For KIC002697388, we chose regions that appeared
pulsation-free between 50-75, and 350-450~$\mu$Hz for the low-frequency
regime with a limit of ranging from 0.23~mma on the long side to 0.13~mma
on the short side and
2000-3700~$\mu$Hz for the high-frequency regime with a detection
limit of 0.11~mma.
For KIC003527751 we used low-frequency regions of 20-80 and 300-360~$\mu$Hz with
resulting limits of 0.11 and 0.07~mma, respectively, 
and 2000-3600$\mu$Hz for a high-frequency
range with a limit of 0.07~mma.
The regions used to calculate the noise were: 20-75, 450-500
and 2000-3400~$\mu$Hz for KPD1943 which produced a long-period detection
limit of 0.09~mma and a short-period limit of 0.07~mma.

In the tables, we include a signal-to-noise (S/N) measurement for each frequency,
measured in standard deviations with the detection limit shown as a solid
(blue) horizontal line in the figures. Also in the figures, we include another
more conservative estimate which uses 
the mean of the peaks in the temporal spectra, rather 
than the mean of the entire spectra shown as a red line. 

\subsection{V1093~Her variables}
For the two V1093~Her variables, the pulsation amplitudes
are readily seen in the temporal spectra. Because of their
large contamination fractions, it is expected that compared to ``normal''
V1093~Her variables, these all have some high-amplitude frequencies.
Additionally, they are all clearly multimode pulsators with many peaks
in their temporal spectra. NLLS fitting and prewhitening was a pretty
straightforward exercise, except where noted below for individual
stars.

\subsubsection{KIC007664467}
\begin{figure}
 \psfig{figure=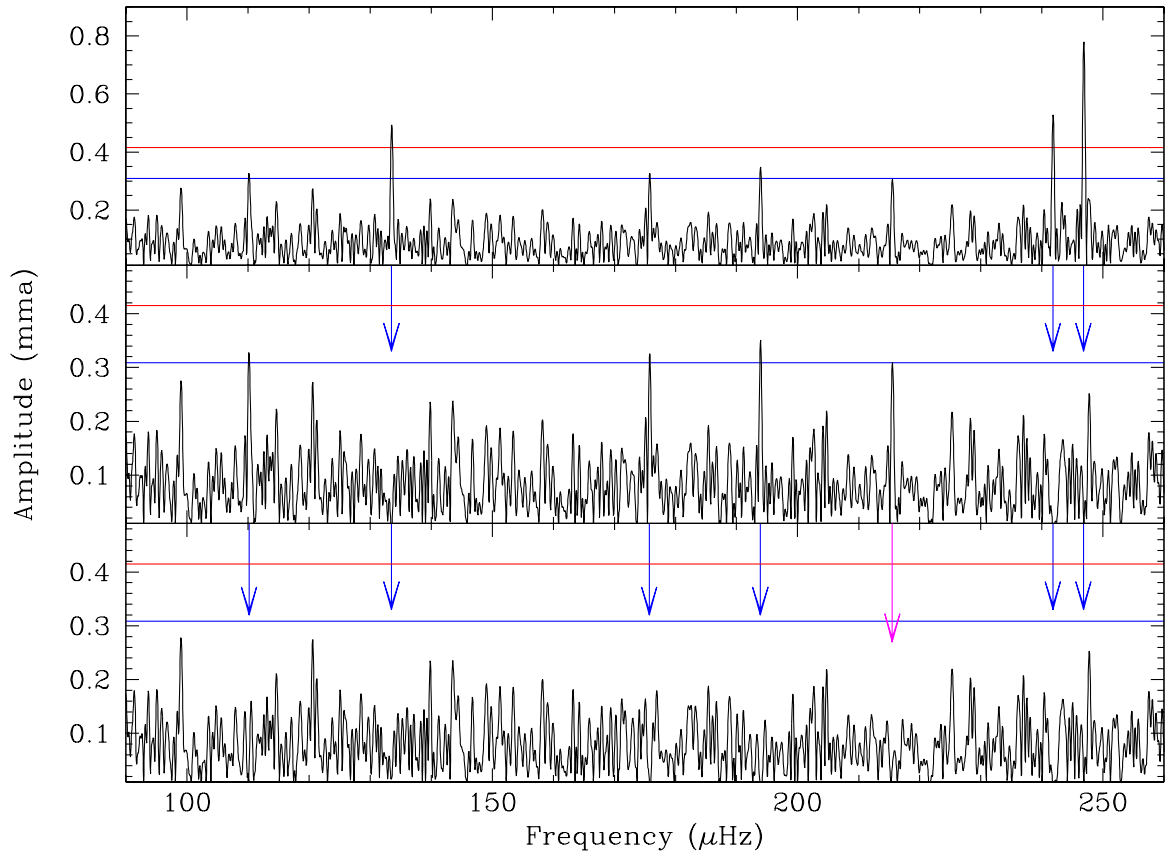,angle=-90,width=\textwidth}
\caption{Temporal spectrum and prewhitening sequence for KIC007664467.
The solid (blue) horizontal 
line is the $4\sigma$ detection limit, the dashed  (red) horizontal line is the 
peak $4\sigma$ detection limit, and arrows indicate prewhitened
frequencies.} \label{fig05a}
\end{figure}

KIC007664467 ($K_p=16.45,\,
T_{\rm eff}\approx 26\,800\pm 500\,$K$,\, \log g=5.17\pm 0.08$; Paper I) 
has the lowest S/N peaks of all
the stars in this paper, resulting in the fewest frequencies detected
(six with confidence, another one less-so at a S/N of 3.76).
These are listed in Table~\ref{tab02} along with their amplitudes and
periods and
indicated in Fig.~\ref{fig05a} with blue and magenta
arrows, respectively. Prewhitening effectively removes all power above
the detection limit and clearly
there are other peaks below the detection limit which are likely produced
by stellar variations. With improved data, which we anticipate obtaining
in the future with \emph{Kepler}, we should be able to ascertain if
those peaks are intrinsic to the star or noise.

\begin{table}
\caption{Frequencies, periods, amplitudes, and S/N for KIC007664467.
Frequencies below the $4\sigma$ detection limit
are listed as \emph{suggested}. \label{tab02} }
\begin{tabular}{lcccc}
\hline
ID & Frequency & Period & Amplitude &  S/N\\
 & ($\mu$Hz) & (s) & (mma)  &  $\sigma$\\ \hline
f1 & 110.179 (0.045)  &       9076.175 (3.719)  & 0.317 (0.060) & 4.1 \\
f2 & 133.551 (0.030)  &       7487.792 (1.683)  & 0.478 (0.060) & 6.2 \\
f3 & 175.789 (0.044)  &       5688.625 (1.423)  & 0.326 (0.060) & 4.2 \\
f4 & 193.932 (0.041)  &       5156.458 (1.081)  & 0.352 (0.060) & 4.6 \\
f5 & 241.826 (0.028)  &       4135.197 (0.486)   & 0.505 (0.060) & 6.5 \\
f6 & 246.872 (0.019)  &       4050.680 (0.308)  & 0.765 (0.060) & 9.9 \\
 & & & & \\
\hline
\multicolumn{4}{c}{Suggested frequency} &  \\
s7 & 215.491 (0.050)  &       4640.562 (1.081)  &  0.285 (0.060) & 3.7 \\
\hline
\end{tabular}
\end{table}

\subsubsection{KIC010670103}
\begin{figure}
 \psfig{figure=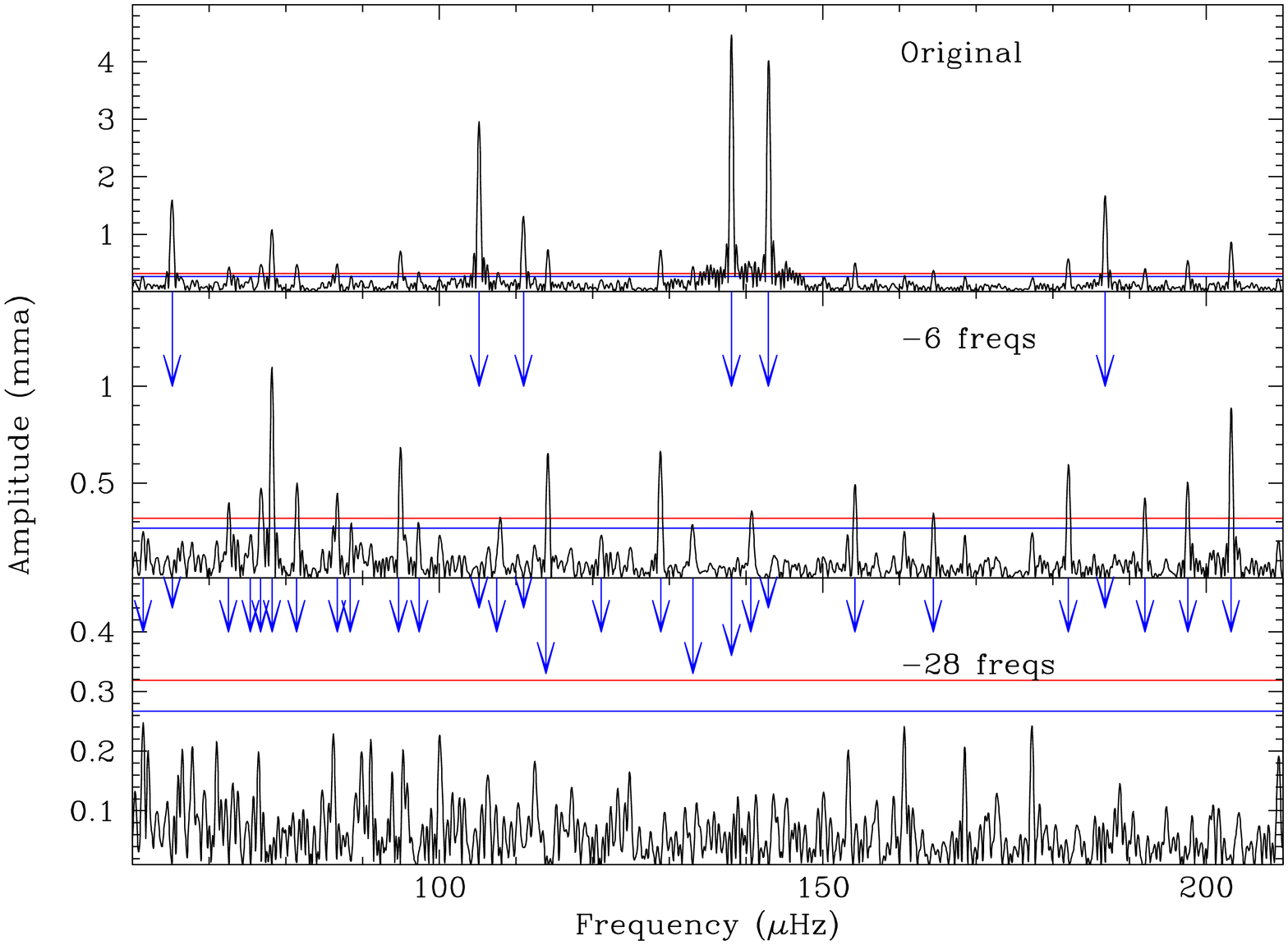,angle=-90,width=\textwidth}
\caption{Same as Fig.~\ref{fig05a} for KIC010670103. The middle and bottom
panels are prewhitened by 6 and 28 frequencies, respectively.}
\label{fig06}
\end{figure}

We detected a total of 28 frequencies for KIC010670103 
($K_p=16.53,\, T_{\rm eff}\approx 20\,900\pm 300\,$K$,\, \log g=5.11\pm 0.04$; Paper I), 
with lots of power still remaining, but just below the detection limit. 
These frequencies 
are listed in Table~\ref{tab03} along with their amplitudes and periods and
indicated in Fig.~\ref{fig06} by arrows.

\begin{table}
\caption{Frequencies, periods, amplitudes, and S/N for KIC010670103.
Possible combination and difference frequencies are marked as $^C$ and
$^D$ and a rudimentary fit of the form $P_{\ell}=P_{\ell o}+n\cdot\Delta 
P_{\ell}$ is given in columns 4 and 5 and discussed in \S4.1. Column
6 provides the difference between the observed and asymptotic frequencies.
Formal least-squares errors are in parentheses.\label{tab03}}
\begin{tabular}{lccccccc}
\hline
ID & Frequency & Period  & $\ell$ & n & $\Delta\nu$ & Amplitude & S/N  \\
 & ($\mu$Hz) & (s) &  & & ($\mu$Hz) &(mma)  &  $\sigma$\\ \hline
f1$^D$& 61.387 (0.050) & 16290.060 (13.247)& 1 & 31 & 0.03 & 0.246 (0.051)& 4.2 \\
f2$^D$& 65.173 (0.008) & 15343.676 (1.790) & - & - & - &1.614 (0.051) & 27.2 \\
f3 & 72.571 (0.031) & 13779.639 (5.790) & 1 & 21 & 0.04 &0.409 (0.051) &  6.9 \\
f4 & 75.399 (0.051) & 13262.826 (8.883) & 1 & 19 & 0.12 &0.251 (0.052) &  4.2 \\
f5 & 76.807 (0.030) & 13019.718 (5.072) & 1 & 18 & 0.08 &0.423 (0.051) &  7.1 \\
f6 & 78.197 (0.012) & 12788.238 (1.970) & 1 & 17 & 0.04 &1.063 (0.051) & 17.9 \\
f7 & 81.445 (0.030) & 12278.230 (4.476) & 1 & 15 & 0.01 &0.425 (0.051) &  7.2 \\
f8 & 86.702 (0.029) & 11533.722 (3.872) & 1 & 12 & 0.06 &0.426 (0.051) &  7.2 \\
f9 & 88.368 (0.048) & 11316.349 (6.162) & 1 & 11 & 0.32 &0.258 (0.051) &  4.4 \\
f10 & 94.942 (0.019) & 10532.747 (2.096) & 1 &  8 & 0.10 &0.661 (0.051) & 11.2 \\
f11 & 97.361 (0.042) & 10271.031 (4.400) & 1 &  7 & 0.01 &0.298 (0.051) &  5.0 \\
f12 & 105.195 (0.004) &9506.147 (0.380) & 1 &  4 & 0.12 &2.984 (0.051)  & 50.4 \\
f13 & 107.924 (0.035) &9265.798 (3.012) & 1 &  3 & 0.01 &0.362 (0.051)  &  6.1 \\
f14 & 110.956 (0.010) &9012.574 (0.823) & 1 &  2 & 0.03 &1.255 (0.051)  & 21.2 \\
f15 & 114.171 (0.019) &8758.812 (1.433) & 1 &  1 & 0.06 &0.672 (0.051)  & 11.3 \\
f16 & 121.073 (0.044) &8259.471 (3.007) & 1 & -1 & 0.02 &0.247 (0.045)  &  4.2 \\
f17 & 128.838 (0.016) &7761.695 (0.973) & 1 & -3 & 0.05 &0.678 (0.045)  & 11.4 \\
f18 & 133.053 (0.037) &7515.790 (2.062) & 1 & -4 & 0.15 &0.300 (0.045)  &  5.1 \\
f19 & 138.099 (0.002) &7241.182 (0.130) & 2 & 16& 0.10 &4.456 (0.046)  & 75.2 \\
f20 & 140.711 (0.030) &7106.778 (1.498) & 2 & 15& 0.12 &0.375 (0.046)  &  6.3 \\
f21 & 142.938 (0.003) &6996.041 (0.133) & 1 & -6 & 0.19 &4.063 (0.045)  & 68.6 \\
f22$^C$& 154.211 (0.022) &6484.639 (0.925) & 1 & -8 & 0.43 &0.508 (0.047)  &  8.6 \\
f23$^C$& 164.421 (0.030) &6081.943 (1.092) & 2 & 8 & 0.01 &0.379 (0.047)  &  6.4 \\
f24$^C$& 182.038 (0.019) &5493.347 (0.575) & 2 & 4 & 0.19 &0.588 (0.047)  &  9.9 \\
f25$^C$& 186.807 (0.007) &5353.106 (0.192)  & 2 & 3 & 0.02&1.675 (0.047)  & 28.3 \\
f26$^C$& 191.993 (0.026) &5208.532 (0.701)  & 2 & 2 & 0.02&0.433 (0.047)  &  7.3 \\
f27$^C$& 197.565 (0.022) &5061.637 (0.563)  & 2 & 1 & 0.03&0.509 (0.047)  &  8.6 \\
f28$^C$& 203.248 (0.013) &4920.108 (0.314) & 2 & 0 & 0.13 &0.863 (0.047)  & 14.6 \\
\hline
\end{tabular}
\end{table}

As indicated in Table~\ref{tab03}, several of the frequencies appear
as possible combinations or differences
of other frequencies, some with multiple combinations which could
interact in very complex ways. As model
frequency densities are quite high (see Paper IV), it is likely these are
chance alignments and so are mentioned solely for completeness. A far
more compelling interpretation of the frequency structure appears in \S4.1.
As examples of possible complex interactions, f1 is a 
combination of three different frequency differences, the most obvious of 
which are differences involving high-amplitude peaks: f19-f5 and f21-f7,
but it is also a combination of f24-f16. 
f2 is a complex combination with some high-amplitude peaks (f19-f3,
f21-f7) and some lower-amplitude ones (f20-f4, f25-f16, and f28-f19);
although some of these lower-amplitude frequencies are themselves
possible combinations of other frequencies. KIC002697388, KIC003527751,
and KPD~1943 are marked in a similar way in Tables~\ref{tab04}, \ref{tab05},
and \ref{tab01}, though we will not discuss
their combinations in the text as we feel these are most likely
chance superpositions caused by a high frequency density spanning 
a large frequency range.

\subsection{Hybrid variables}
For these stars, the pulsations seem to break into three regions,
just as for BA09 \citep{andy09}, with periodicities longer
than about 40 minutes, typical for $g-$mode pulsators, periodicities
shorter than five minutes, typical of $p-$mode pulsators, and then
a group that are in between with periods ranging from 15 to 30 
minutes.

\subsubsection{KIC002697388}
\begin{figure}
 \psfig{figure=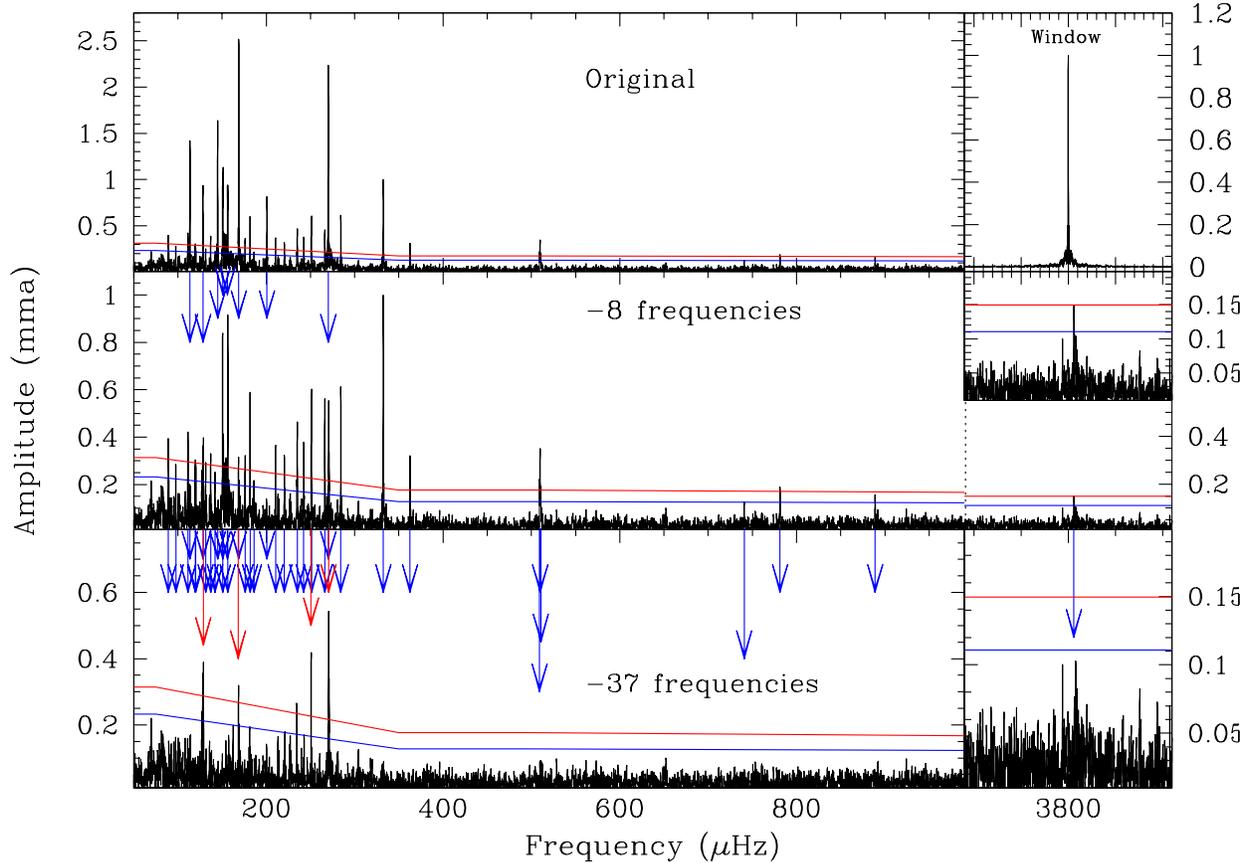,angle=-90,width=\textwidth}
\caption{Same as Fig.~\ref{fig05a} for KIC002697388. The 
middle panel has eight frequencies removed and the
bottom panel is prewhitened by 37 frequencies. Red arrows indicate
likely frequencies that were not fitted (see Fig.~\ref{fig08} for an
enlarged view).
Note that the vertical scale changes with each panel.
The right panels show the window function on top, the middle panel is
broken into two sub-panels so the high-frequency regime can be
plotted on the same vertical scale as the left panel (separated by a
dashed blue line) and an enlarged view better shows the peak in the
original FT. The bottom panel is the prewhitened FT.} \label{fig07}
\end{figure}

\begin{figure}
 \psfig{figure=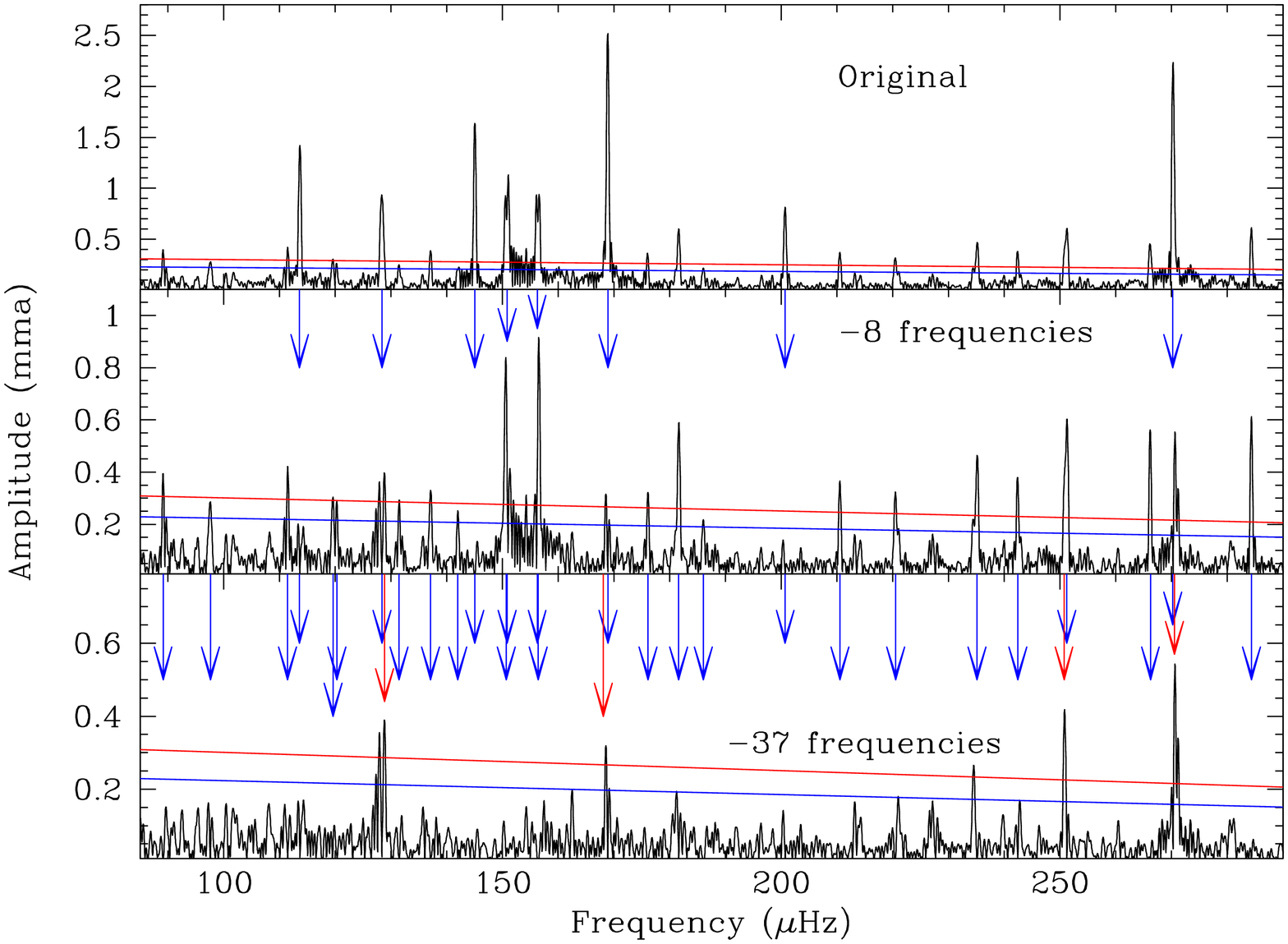,angle=-90,width=\textwidth}
\caption{Temporal spectrum of KIC002697388  between 100 and $300\,\mu$Hz.
Note that the vertical scale changes with each panel.} \label{fig08}
\end{figure}

 In total, we detected
37 long-period pulsations leaving four regions with
unresolved frequencies in KIC002697388 ($K_p=15.39,\, T_{\rm eff}\approx 23\,900\pm
300\,$K$,\,\log g=5.32\pm 0.03$; Paper I).
These are listed in Table~\ref{tab04} along with their amplitudes and
periods and
indicated in Fig.~\ref{fig07} with 
arrows. Unresolved power remains in the FT above the
detection limit, which could not be NLLS fitted 
and the most obvious of these are indicated by longer (red)
arrows. Figure~\ref{fig08} is an enlarged view of the low-frequency
region.

We also detected a high-frequency peak above the
$4\sigma$ limit that was not an artefact caused by long-cadence readout.
The peak is low amplitude but a $5.4\sigma$ detection in the combined data
which is significant above the noise. The hybrids are discussed further
in \S4.3 and until KIC002697388 receives further observations (Q5),
we consider it as a candidate hybrid pulsator. 

\begin{table}
\caption{Frequencies, periods, amplitudes, and S/N for KIC002697388
Formal least-squares errors are in parentheses. $^C$
and $^D$ indicates possible combination and difference 
frequencies. Unresolved frequencies which could not be NLLS fitted are listed
at the bottom with amplitudes estimated directly from the FT.\label{tab04}}
\begin{tabular}{lcccc}
\hline
ID & Frequency & Period  & Amplitude & S/N \\
 & ($\mu$Hz) & (s) & (mma) & \\ \hline
f1$^D$ & 89.109 (0.017)  &11222.207 (2.079)& 0.385 (0.026) & 6.8 \\
f2 & 97.579 (0.023)  &10248.154 (2.432)& 0.274 (0.026) & 4.9 \\
f3 & 111.477 (0.015) &8970.500 (1.205) &0.426 (0.026) & 7.8 \\
f4$^D$ & 113.649 (0.006) &8799.008 (0.459) &1.441 (0.036) & 26.5 \\
f5 & 119.504 (0.024) &8367.893 (1.699) &0.272 (0.027) & 5.0 \\
f6 & 120.305 (0.027) &8312.182 (1.867) &0.244 (0.027) & 4.5\\
f7 & 128.350 (0.009) &7791.189 (0.548) &0.945 (0.036) &17.8\\
f8 & 131.442 (0.022) &7607.932 (1.267) &0.291 (0.026) & 5.5\\
f9 & 137.085 (0.020) &7294.763 (1.058) &0.320 (0.026) &6.1\\
f10 & 141.972 (0.028) &7043.651 (1.375) &0.229 (0.026) &4.4\\
f11 & 145.026 (0.005) &6895.323 (0.253) &1.609 (0.036) &31.2\\
f12 & 150.670 (0.027) &6637.021 (1.178) &1.371 (0.259) &26.9\\
f13 & 150.891 (0.025) &6627.321 (1.090) &1.480 (0.259) &29.0\\
f14 & 156.278 (0.044) &6398.864 (1.813) &2.557 (1.765) &50.7\\
f15 & 156.397 (0.044) &6393.989 (1.813) &2.553 (1.765) &50.6\\
f16 & 168.861 (0.003) &5922.022 (0.118) &2.529 (0.036) &51.3\\
f17$^D$ & 176.073 (0.019) &5679.474 (0.601) &0.336 (0.026) &6.9\\
f18$^D$ & 181.607 (0.011) &5506.391 (0.323) &0.591 (0.026) &12.3\\
f19 & 186.048 (0.027) &5374.965 (0.769) &0.236 (0.026) &4.9\\
f20 & 200.679 (0.010) &4983.073 (0.257) &0.825 (0.036) &17.8\\
f21 & 210.502 (0.017) &4750.550 (0.387) &0.365 (0.026) &8.0\\
f22 & 220.453 (0.020) &4536.116 (0.407) &0.317 (0.026) &7.1\\
f23 & 235.124 (0.014) &4253.082 (0.250) &0.453 (0.026) &10.5\\
f24 & 242.388 (0.016) &4125.623 (0.268) &0.398 (0.026) &9.4\\
f25 & 251.250 (0.010) &3980.103 (0.163) &0.609 (0.026) &14.6\\
f26 & 266.151 (0.011) &3757.261 (0.160) &0.553 (0.026) &13.7\\
f27 & 270.254 (0.004) &3700.224 (0.052) &2.230 (0.036) &55.8\\
f28$^C$ & 284.316 (0.010) &3517.219 (0.128) &0.606 (0.026) &15.7\\
f29 & 332.366 (0.006) &3008.732 (0.054) &1.000 (0.025) &29.2\\
f30$^C$ & 362.698 (0.019) &2757.118 (0.141) &0.323 (0.025) &10.1\\
f31$^C$ & 509.169 (0.039) &1963.985 (0.149) &0.165 (0.026) &5.2\\
f32$^C$ & 509.947 (0.019) &1960.990 (0.074) &0.352 (0.027) &11.0\\
f33$^C$ & 510.668 (0.031) &1958.219 (0.120) &0.212 (0.026) &6.6\\
f34$^C$ & 740.877 (0.047) &1349.752 (0.086) &0.126 (0.025) &4.0\\
f35$^C$ & 781.382 (0.031) &1279.784 (0.051) &0.192 (0.025) &6.0\\
f36 & 888.917 (0.038) &1124.964 (0.048) &0.157 (0.025) &4.9\\
f37 & 3805.906 (0.040)& 262.750 (0.003) &0.147 (0.024)  &5.3\\
 & & & & \\
\hline
\multicolumn{4}{c}{Unresolved frequencies}\\ 
u38 & 128.8 &  & 0.39 & \\
u39 & 168.1 &  & 0.32 & \\
u40 & 250.7 &  & 0.42 & \\
u41$^C$ & 270.5 &  & 0.55 & \\ \hline
\end{tabular}
\end{table}

As observed for  BA09 \citep{andy09}, several of the 
intermediate frequencies are possible combination frequencies. As
this frequency region is less densely populated in models, 
it is likely some of these are real combinations.
However, not all of the 
frequencies in this range can be attributed to combination frequencies.

\subsubsection{KIC003527751}
\begin{figure}
 \psfig{figure=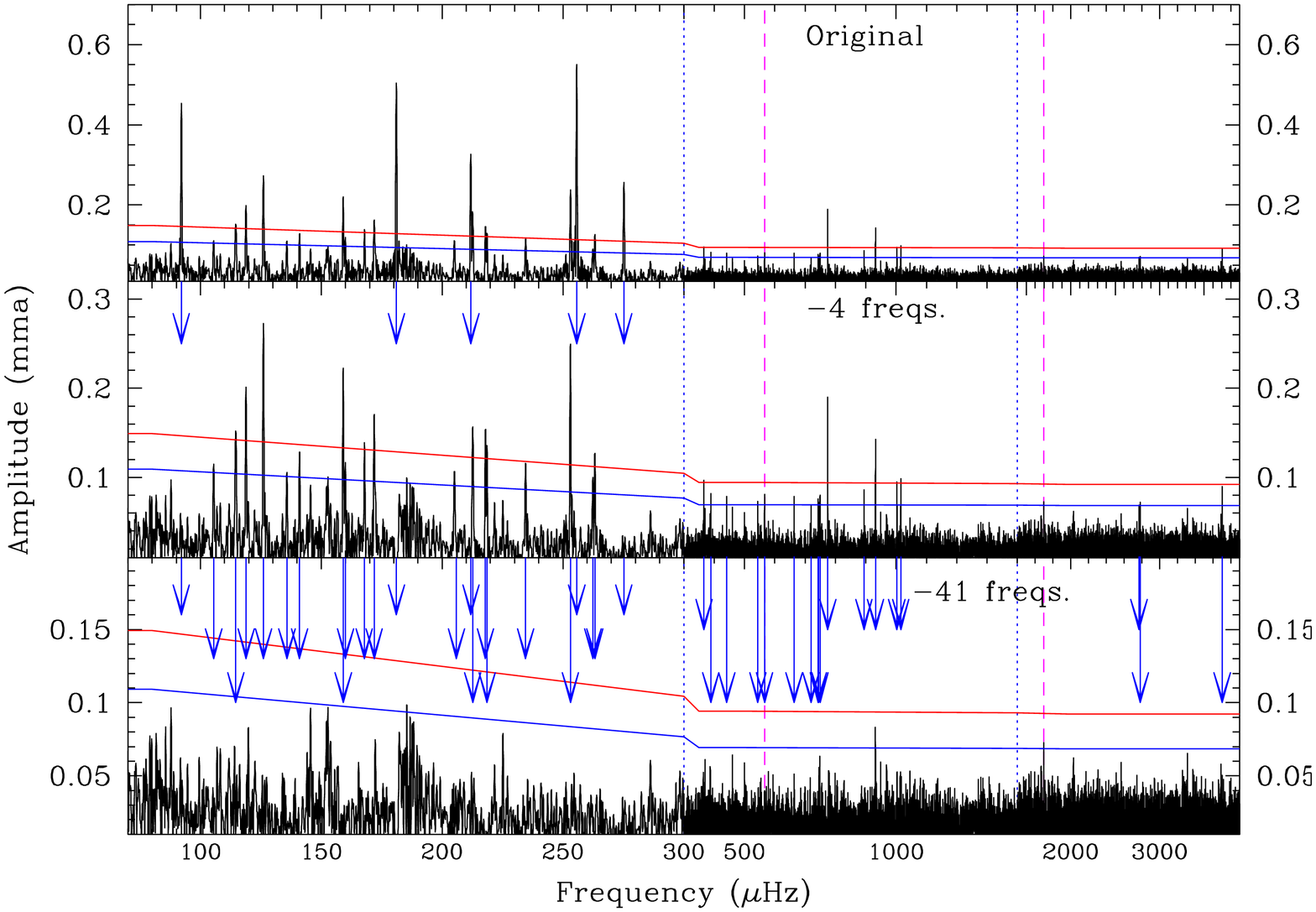,angle=-90,width=\textwidth}
\caption{Same as Fig.~\ref{fig05a} for KIC003527751. The frequency axis
is continuous across the figure, but changes scale at the
dotted vertical lines. The top panel shows the original spectrum with
subsequent panels prewhitened by four and 41 frequencies. The top
right panel shows the window function for these data.
Note that the vertical scale changes with each panel.} \label{fig09}
\end{figure}

All of the pulsation amplitudes of KIC003527751 ($K_p=14.86$, 
$T_{\rm eff}\approx 27\,900\pm 200\,$K$,\,\log g=5.37\pm 0.09$; Paper I) 
are quite low and would
almost certainly be missed in ground-based observations, yet they are
easily detected in these \emph{Kepler} data. 
In total, we detected
41 pulsation frequencies 
with no residuals above the detection limit.
These are listed in Table~\ref{tab05} along with their amplitudes and
periods and indicated in Fig.~\ref{fig09} with arrows.

\begin{table}
\caption{Frequencies, periods, amplitudes, and S/N for KIC003527751.
Possible combination and difference frequencies are noted with $^C$
and $^D$ in column 1.
Formal least-squares errors are in parentheses.\label{tab05}}
\begin{tabular}{lcccc}
\hline
ID & Frequency & Period  & Amplitude & S/N\\
 & ($\mu$Hz) & (s) & (mma) & $\sigma$\\ \hline
f1 & 92.144 (0.008) &10852.611 (0.962) &0.455 (0.016) & 17.0\\
f2 & 105.478 (0.030)& 9480.648 (2.717) &0.116 (0.015) & 4.4\\
f3 & 114.625 (0.023) &8724.069 (1.752) &0.153 (0.015) &6.0\\
f4 & 118.825 (0.017) &8415.768 (1.207) &0.207 (0.015) &8.1\\
f5 & 126.070 (0.013) &7932.131 (0.821) &0.269 (0.015) &10.7\\
f6 & 135.767 (0.034) &7365.565 (1.833) &0.104 (0.015) &4.2\\
f7$^D$ & 141.016 (0.027) &7091.397 (1.380) &0.128 (0.015) &5.2\\
f8 & 159.098 (0.016) &6285.436 (0.615) &0.233 (0.015) &9.8\\
f9 & 159.878 (0.028) &6254.758 (1.080) &0.132 (0.015) &5.6\\
f10 & 167.871 (0.024) &5956.947 (0.845) &0.147 (0.015) &6.3\\
f11 & 171.928 (0.020) &5816.402 (0.675) &0.176 (0.015) &7.6\\
f12 & 181.052 (0.007) &5523.277 (0.225) &0.503 (0.016) &22.2\\
f13 & 205.104 (0.031) &4875.586 (0.743) &0.112 (0.015) &5.2\\
f14 & 211.872 (0.011) &4719.823 (0.251) &0.332 (0.016) &15.6\\
f15 & 212.672 (0.022) &4702.076 (0.489) &0.158 (0.015) &7.4\\
f16$^D$ & 217.937 (0.026) &4588.472 (0.557) &0.145 (0.016) &6.9\\
f17$^D$ & 218.584 (0.031) &4574.895 (0.649) &0.124 (0.016) &5.9\\
f18$^D$ & 234.546 (0.031) &4263.556 (0.569) &0.119 (0.016) &5.9\\
f19 & 253.095 (0.015) &3951.085 (0.230) &0.257 (0.016) &13.2\\
f20 & 255.694 (0.007) &3910.932 (0.103) &0.562 (0.016) &29.1\\
f21$^C$& 262.376 (0.041) &3811.326 (0.596) &0.094 (0.016) &4.9\\
f22 & 263.181 (0.030) &3799.665 (0.436) &0.128 (0.016) &6.7\\
f23 & 275.216 (0.015) &3633.516 (0.196) &0.251 (0.016) &13.6\\
f24$^C$& 366.437 (0.035) &2728.982 (0.264) &0.097 (0.014) &5.7\\
f25$^C$ & 388.851 (0.042) &2571.681 (0.278) &0.081 (0.014) &4.7\\
f26$^C$ & 441.541 (0.043) &2264.793 (0.222) &0.079 (0.014) &4.6\\
f27$^C$ & 543.979 (0.048) &1838.307 (0.163) &0.071 (0.014) &4.1\\
f28 & 566.404 (0.043) &1765.523 (0.133) &0.080 (0.014) &4.7\\
f29$^C$ & 663.714 (0.044) &1506.672 (0.099) &0.078 (0.014) &4.6\\
f30 & 719.535 (0.049) &1389.787 (0.095) &0.069 (0.014) &4.0\\
f31 & 743.649 (0.098) &1344.720 (0.177) &0.098 (0.049) &5.7\\
f32 & 743.918 (0.105) &1344.234 (0.189) &0.091 (0.049) &5.3\\
f33$^C$& 749.222 (0.042) &1334.717 (0.075) &0.081 (0.014) &4.7\\
f34 & 774.700 (0.018) &1290.822 (0.030) &0.190 (0.014) &11.1\\
f35 & 894.390 (0.039) &1118.081 (0.049) &0.087 (0.014) &5.1\\
f36$^C$ & 932.440 (0.024) &1072.455 (0.028) &0.141 (0.014) &8.2\\
f37$^C$ & 1002.470 (0.036)& 997.536 (0.036) &0.095 (0.014) &5.5\\
f38$^C$ & 1015.819 (0.035)& 984.427 (0.034) &0.098 (0.014) &5.7\\
f39 & 2767.427 (0.050)& 361.346 (0.007) &0.068 (0.014) &4.0\\
f40 & 2782.505 (0.048)& 359.388 (0.006) &0.071 (0.014) &4.1\\
f41 & 3703.293 (0.038)& 270.030 (0.003) &0.089 (0.014) &5.2\\
\hline
\end{tabular}
\end{table}

\subsubsection{KPD~1943 (KIC005807616)}
\begin{figure}
\psfig{figure=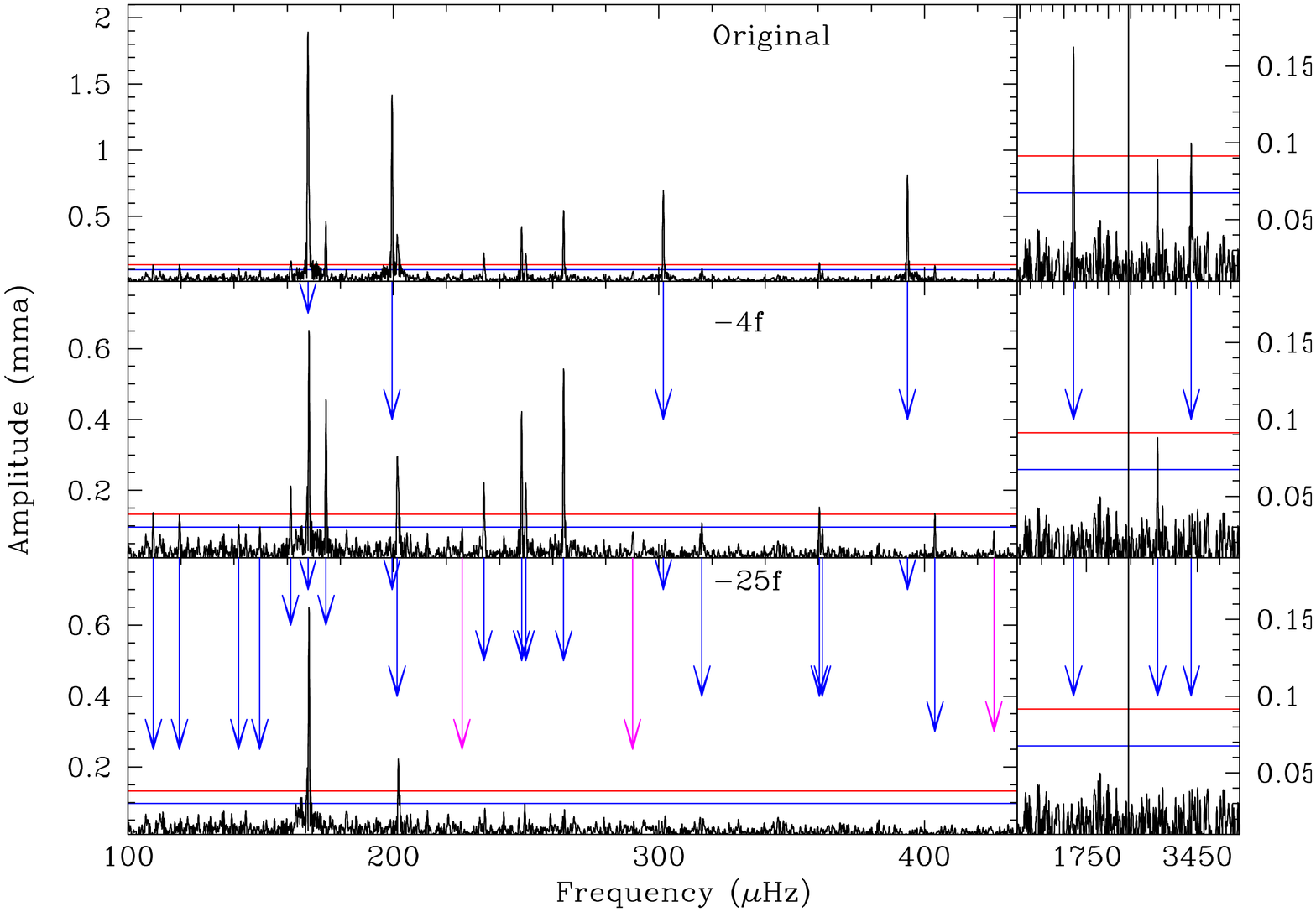,angle=-90,width=\textwidth}
\caption{Same as Fig.~\ref{fig05a} for KPD~1943.
 The middle panel is prewhitened by 4 frequencies and the bottom is
panel is prewhitened by 25 frequencies. Unresolved regions are shown
in Fig.~\ref{fig04}.}
\label{fig03}
\end{figure}

In total, we detected
21 frequencies with confidence,
four frequencies that were marginally below the detection limit,
and still there remained three regions of
unresolved power for KPD~1943 ($K_p=15.02,\,T_{\rm eff}\approx 27\,100\pm
200\,$K$,\,\log g=5.51\pm 0.02$; Paper I).
These are listed in Table~\ref{tab01} along with their amplitudes and
periods and
indicated in Fig.~\ref{fig03} with blue and magenta
arrows, respectively. We note that an independent analysis in Paper~IV obtained
a slightly different, but generally consistent, set of frequencies.

\begin{figure}
 \psfig{figure=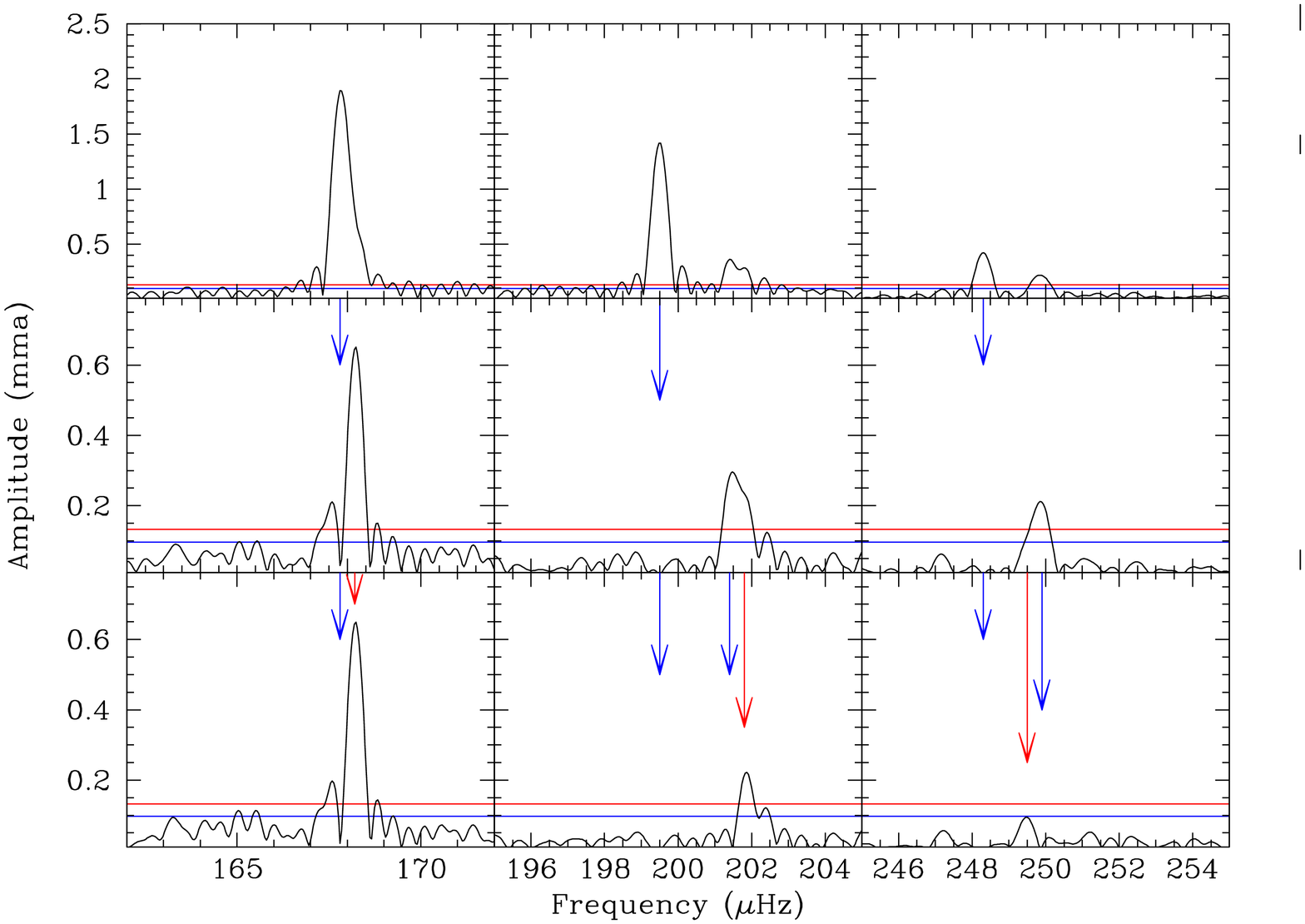,angle=-90,width=\textwidth}
\caption{Temporal spectra for KPD~1943 showing unresolved regions. Each panel
is $10\mu$Hz wide and blue arrows indicate prewhitened frequencies while 
red ones indicate unresolved frequencies listed in Table~\ref{tab01}.} 
\label{fig04}
\end{figure}

Figure~\ref{fig04} shows the unresolved regions, with red arrows indicating
the frequencies listed in Table~\ref{tab01}.

\begin{table}
\caption{Frequencies, periods, amplitudes, and S/N for KPD~1943.
Formal least-squares errors are in parentheses. Possible
combination and difference frequencies are marked in column 1 with $^C$
and $^D$.\label{tab01}}
\begin{tabular}{lcccc}
\hline
ID & Frequency & Period  & Amplitude & S/N \\
 & ($\mu$Hz) & (s) & (mma) & S/N \\ \hline
f1$^D$& 109.50 (0.04)     &   9132.311 (2.979)   &    0.14 (0.02) &5.8 \\
f2 & 119.44 (0.04)     &   8372.266 (2.611)   &    0.13 (0.02) &5.4 \\
f3 & 141.73 (0.03)     &   7055.779 (1.667)   &    0.10 (0.01) &4.1 \\
f4 & 149.73 (0.05)     &   6678.527 (2.251)   &    0.10 (0.02) &4.1 \\
f5$^D$& 161.32 (0.02)     &   6198.841 (0.892)   &    0.21 (0.02) &8.7 \\
f6 & 167.82 (0.00)     &   5958.808 (0.092)   &    1.90 (0.02) &78.4 \\
f7 & 174.64 (0.01)     &   5726.085 (0.353)   &    0.46 (0.02) &19.0 \\
f8 & 199.50 (0.00)     &   5012.561 (0.088)   &    1.41 (0.02) &58.2 \\
f9 & 201.49 (0.02)     &   4963.086 (0.410)   &    0.30 (0.02) &12.4 \\
f10 & 234.06 (0.02)    &   4272.329 (0.407)   &    0.22 (0.02) &9.1 \\
f11 & 248.30 (0.01)    &   4027.352 (0.191)   &    0.42 (0.02) &17.3 \\
f12 & 249.87 (0.02)    &   4002.153 (0.374)   &    0.21 (0.02) &8.7 \\
f13 & 264.11 (0.01)    &   3786.345 (0.130)   &    0.54 (0.02) &22.3 \\
f14 & 301.67 (0.01)    &   3314.876 (0.078)   &    0.69 (0.02) &28.5 \\
f15$^C$ & 316.18 (0.05)    &   3162.719 (0.463)   &    0.11 (0.02) &4.5 \\
f16 & 360.44 (0.03)    &   2774.381 (0.258)   &    0.15 (0.02) &6.2 \\
f17 & 393.61 (0.01)    &   2540.590 (0.039)   &    0.82 (0.02) &33.8 \\
f18 & 403.96 (0.04)    &   2475.513 (0.222)   &    0.13 (0.02) &5.4 \\
f19 & 1744.29 (0.02)   &   573.300 (0.007)    &    0.16 (0.01) &9.5 \\
f20 & 3432.08 (0.04)   &   291.369 (0.003)    &    0.09 (0.01) &5.3 \\
f21 & 3447.23 (0.04)   &   290.088 (0.003)    &    0.10 (0.01) &5.9 \\ \hline
\multicolumn{4}{c}{Suggested Frequencies} \\
s22$^D$ & 225.90 (0.05)   &    4426.714 (0.964)   &    0.10 (0.02) &4.1\\
s23 & 290.19 (0.06)   &    3446.011 (0.720)   &    0.08 (0.02) &3.3\\
s24 & 361.63 (0.06)   &    2765.289 (0.443)   &    0.09 (0.02) &3.7\\
s25$^C$& 426.18 (0.06)   &    2346.420 (0.328)   &    0.08 (0.02) &3.3\\ \hline
\multicolumn{4}{c}{Unresolved Frequencies} \\
u26 & 168.2  &  & 0.99 & \\
u27$^D$ & 202.4  &  & 0.36  &\\
u28 & 249.5  &  & 0.16 & \\
\hline
\end{tabular}
\end{table}

\section{Discussion}
The V1093~Her pulsators all have periods longer then an hour and 
the extremes scale roughly with $\log g$ and $T_{\rm eff}$. KIC010670103
is the coolest sdBV detected to date and has the longest periods, reaching
nearly 4.5~h in duration. Such long periods would obviously be
very difficult to detect from Earth with interrupting diurnal cycles.
KIC007664467 has the fewest frequencies detected in this sample and
they are easily resolved. However, this is likely based on the detection
limit, so with more data, we will likely find more frequencies.
KIC010670103 is the richest V1093~Her pulsator known to date with 28
frequencies.

The hybrid pulsators discovered by \emph{Kepler} are also extremely rich
pulsators with over 100 frequencies detected between the three stars. The
$p-$mode pulsations were a surprise discovery. While two of the hybrids 
have temperatures and gravities similar to non-\emph{Kepler} hybrids,
we did not anticipate hybrids where the $p-$mode amplitudes were 
smaller than the $g-$mode ones. Similarly, the candidate hybrid
KIC002697388 is surprisingly cool. Should subsequent \emph{Kepler}
observations (during Q5) confirm the hybrid nature of KIC002697388,
it would indeed represent an intriguing object. Driving $p-$mode pulsations
at those temperatures is not a problem, but rather depends on the amount
of iron enhancement \citep[see Fig.~2 of ][]{charp07}. Defining the
extent of the $p-$mode instability region constrains the minimum amount
of Z-bump enhancement in the driving region.

\subsection{Mode identifications using asymptotic relations}
While the crowded $g-$mode frequency density may appear as a disadvantage
for model fitting, because we have reached a region where asymptotic
relationships may apply, it could prove useful for mode identifications. In
turn, observational correlations between modes and frequencies can provide
strong model constraints \citep[e.g.,][]{wing91}.

In the asymptotic limit for $n\gg\ell$, $g-$modes should be equally spaced
in period for consecutive values of $n$ \citep{smey}, where $n$ represents
the number of radial nodes and $\ell$ is the number of surface 
nodes. This asymptotic behaviour has been observed very clearly in pulsating
white dwarfs \citep[e.g.][]{kaw94}. The relationship
is  $$\Pi_{n,\ell}=\,n\times
\frac{\Pi_o}{\sqrt{\ell\left(\ell +1\right)}}$$
and indicates a second important feature which is useful for mode
identifications. Modes of consecutively higher degree $\ell$ will
be spaced closer together in a predictable relation
and modes of the same $n$ but differing
$\ell$ will be related in period. In  particular, the relations
between $\ell\,=\,1$ and 2 modes for period spacings and  related overtones
will be $\sqrt{3}$. Specifically, $$\Delta\Pi_{\ell\,=\,2}\,=\,
\frac{\Delta\Pi_{\ell\,=\,1}}{\sqrt{3}}\quad\mbox{ and }\quad\Pi_{n,\ell\,=\,2}\,=\,
\frac{\Pi_{n,\ell\,=\,1}}{\sqrt{3}}$$ for large $n$.

A quick examination of KIC010670103's periods easily shows that there are
common period spacings. As such, we used it as a test to see how thoroughly
asymptotic relations could apply. Additionally,
KIC010670103 has lots of
periods to work with, but not so many as to make the task daunting. As we
do not observe frequency multiplets, we assume that rotational splitting
can be neglected. KIC010670103 shows
several spacings around 145 and 250~s and we also noticed
that the 145~s spacings mostly occurred for shorter periods (higher frequencies)
and the 250~s spacings for longer periods, as would be expected. 
Upon closer examination, it
is obvious that $145\approx 250/\sqrt{3}$ and that several pairs of
periods are also related directly by $1/\sqrt{3}$. This prompted one of
us (sdk) to produce fits to the periods for each spacing of the
form $P_{\ell}=P_{\ell o}+n\cdot\Delta P_{\ell}$. For the 
apparent $\ell =1$ modes, the fit is $P_1=8512.08+n\cdot 251.16$ and for the
$\ell =2$ modes it is $P_2=4916.58+n\cdot 145.59$. The base frequencies
were chosen such that the shortest period in KIC010670103 is $n=0$ and
the associated $\ell =1$ period would have the same $n$ value. This scheme
produces some negative $n$ values for the $\ell =1$ sequence, but it is 
very important to note that while consecutive $n$ values represent
consecutive overtones within the star, the values themselves
are not representative of the actual $n$ value for any given period.
Indeed, they must \emph{not} be associated with the actual value, or
asymptotic relations would not apply. 

We tested the sequence of period spacings using Komogorov-Smirnov statistical tests
\citep{kaw88} and inverse variance tests \citep{od94}, along with Monte Carlo
simulations using randomly selected periods within the observed range.  The observed
sequence is statistically significant at the 99.99\% level based on these tests. 
The results of these fits are shown in columns 4, 5, and 6 of 
Table~\ref{tab03} where the difference between the asymptotic relation
period and the observed period (column 6) is given in frequency, so it can
be compared with the $1/T$ resolution of $0.43\,\mu$Hz. 
Amazingly, these two relations fit 27 of the 28 
periodicities observed in KIC010670103, with f24 fitting at the
$1/T$ resolution limit. Additionally, there are several
pairs of frequencies that are related overtones (same $n$ value). This
provides very powerful evidence that nearly all of the observed
periodicities in KIC010670103 are $\ell\,=\,1$ and 2 modes as the
two sequences are not independent spacings, but rather include
six pairs of periods related by $1/\sqrt{3}$, which means these must
be the same $n$ value (whatever that may be) intrinsically to the star.

Further investigations will be warranted once the longer-duration (Q5
and beyond) data sets are obtained, both for KIC010670103 and other
$g-$mode pulsators. We also note that a satisfactory model fit has
been obtained for KPD~1943 (Paper IV) without the aid of asymptotics
(and indeed the model fit indicates the frequencies are far from 
satisfying the asymptotic behaviour).

\subsection{Short-period amplitude stability}
\begin{figure}
 \psfig{figure=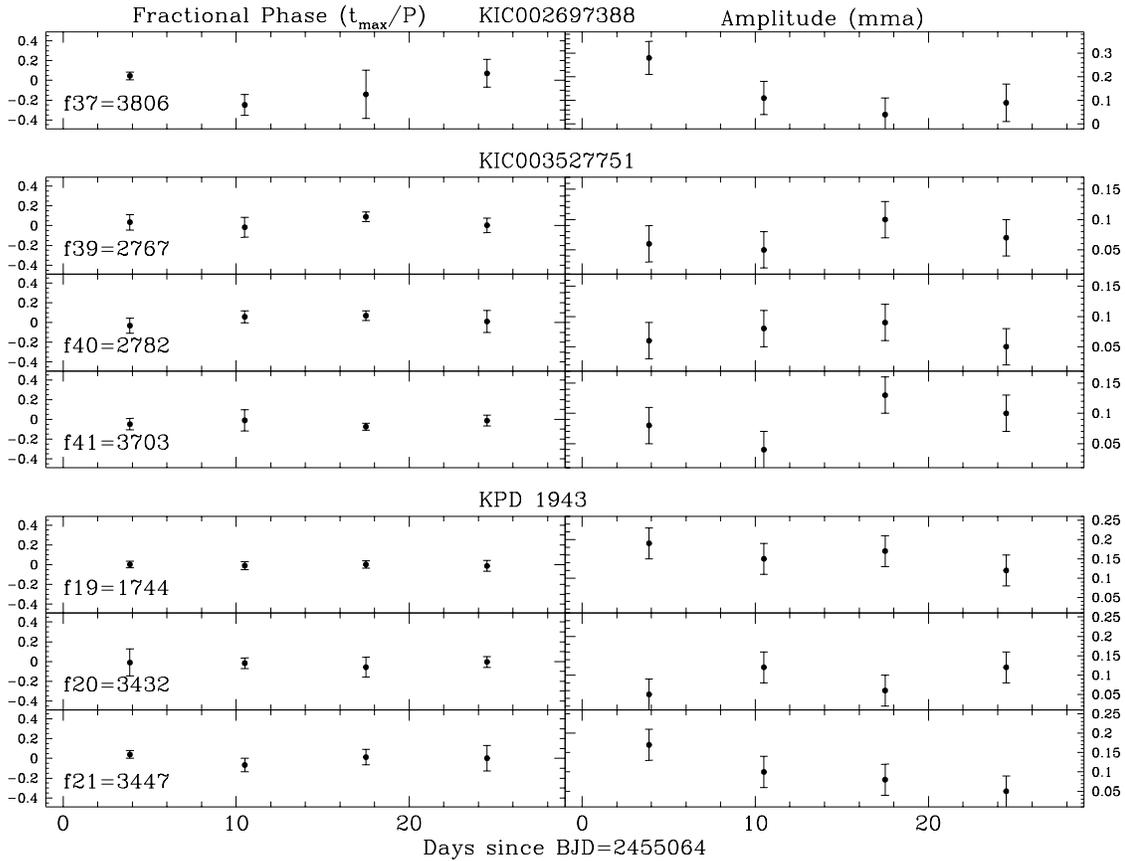,angle=-90,width=\textwidth}
\caption{Phases (left panels) and amplitudes (right panels) for the
high-frequency variations determined over one week intervals.}
\label{fig05b}
\end{figure}

As some low-amplitude short-period $p-$mode frequencies can appear
transitory (Paper I), we divided our data into four one-week sections,
and fixing the frequencies, fitted the phases and amplitudes. As can be seen
in Fig.~\ref{fig05b}, except for KIC002697388, the phases are
stable and while their
amplitudes waver a bit (some get quite low),
nearly all are constant within the $1\sigma$ errorbars. However
while KPD~1943's f21's phases seem stable, the amplitude continually decreases
and the last measurement is nearly undetectable. We also note
that all three of KPD~1943's short-period $p-$mode frequencies lie
within $30\,\mu$Hz of an LC read-time artefact. While we do not think
they are related, it is a coincidence that compels further investigation.
It will be interesting
to see how the high-frequency amplitudes vary, presuming the frequencies 
persist at detectable levels
during \emph{Kepler's} next observational run on these stars. Those data
will also be useful for determining relations to the LC artefacts.

\subsection{Hybrid amplitudes}
Compared to previously known hybrid pulsators, these \emph{Kepler} variables are
unusual in that the
$g-$mode pulsations have higher
amplitudes than the highest $p-$mode ones (or one, for KIC003527751).
For comparison with the previously known hybrids in \S 1,
the highest-amplitude $p-$ and $g-$mode pairs for KIC002697388,
 KIC003527751, and KPD~1930 are
$0.15:2.56$, $0.09:0.56$, and $0.16:1.90$ which 
are reversed from the non-\emph{Kepler} hybrids. While the results 
are very preliminary
and subject to systematics between various spectral line fitting procedures, 
it appears that the $g-$mode dominated hybrids are very slightly cooler 
than the $p-$mode dominated ones. If verified this could indicate a real 
change in how the pulsation power is applied to differing regions of the 
star as $g-$modes are more
sensitive to core conditions while $p-$modes are atmospheric in nature.


\section{Conclusions and future work}
The data presented here confirm the potential of the \emph{Kepler}
mission for sdB asteroseismology. We have identified pulsations with
amplitudes blending into the detection limit. We anticipate that further,
longer-duration observations during \emph{Kepler's} second year of
operation will resolve and detect even more, lower-amplitude frequencies
and unambiguously determine if the first hybrids with lower $p-$ mode
amplitudes have been detected.

Frequency density is not an issue for $g-$mode pulsators \citep{mdr2004_2}
as models provide many closely spaced frequencies in this region
\citep{font,jeff07,hu09}. Yet pulsation models have found it easier to drive
high-degree ($\ell\geq 3$) modes and since the detected amplitudes continue down to the
detection limit, and likely beyond, it
is likely that high-degree modes are present. \emph{Kepler} can test this
critical model assumption in that with more data and a lower detection
threshold, it may be possible to determine if $\ell\geq 3$ are truly
present. As $\ell\geq 3$ have a large degree of
geometric cancellation \citep{charp05a,me1}, if their amplitudes are
intrinsically similar to low-degree modes, then their observed amplitudes
 would be significantly reduced. Sustained \emph{Kepler} photometry will be
able to detect such low-amplitude frequencies.

An asymptotic limit approach has been applied to the period spacings 
of KIC010670103. The results show a good fit to nearly all the observed
periodicities, identifying all but one frequency as $\ell\,=\,1$ or 2
modes. Both the spacings and relations between the two sets agree with the
expected $1/\sqrt{3}$ asymptotic relations to surprising accuracy. This is
remarkable for two reasons: firstly in that it indicates that $\ell\,=\,1$ and
2 modes are not only present, but predominant in at least one star and secondly in
that asymptotic relations may be a useful method for correlating 
periodicities with modes for sdB stars.  A future paper will examine 
the period spacings of all the \emph{Kepler} $g-$mode pulsators and attempt 
to constrain mode identifications.

We have shown results from a study of the first \emph{Kepler} data on 
newly-discovered V1093~Her variables. Both
stars likely have variations of sufficient amplitude to have
been detected from ground-based observations\footnote{With the 
contamination correction, which we did not apply.}, though they are all
fainter than $15^{th}$ magnitude. Yet the majority of
pulsation amplitudes we detect are below 0.1\%, which is very difficult to detect
from the ground, especially taking atmospheric transparency variations 
into consideration. For comparison, power-weighted mean frequencies
were calculated which result in $f_{\rm med}$~=208.0, and 131.6~$\mu$Hz
for KIC007664467, and KIC010670103, respectively.
A ground-based
observing night (of 8~h) would obtain 6.0, and 3.8 continuous
pulsation cycles before having a daytime gap. \emph{Kepler} has obtained
over 550, and 352 pulsation cycles of nearly continuous, 
homogeneously-obtained data. These provide the richest 
$g-$mode pulsation spectra obtained to date, 
which are essential for understanding these complex pulsators.

In these stars we detected seven and 28 frequencies, with the possibility
that some could be combination frequencies.
We have also detected the longest periods (near 4.5~h) known to date, 
for the coolest known sdBV star ($T_{\rm eff}\approx 20\,900\,$K; Paper I),
thus extending the temperature range of pulsators.  As periodicities are
sensitive to the stellar radius, measured via gravities, the stars with the
lowest gravities should have the longest periods. As both stars have similar
gravities, their shortest periods are similar, but KIC010670103's cooler 
nature seems to drive higher overtones (and thus longer periods).

We have examined the first three hybrid subdwarf B pulsators
discovered by the \emph{Kepler} mission. These stars are
unique among hybrid pulsators for several reasons including
the abundance of $g-$mode frequencies, the reversed amplitude
ratios between the $p-$ and $g-$modes compared to non-\emph{Kepler}
hybrids, the complexities of their combination frequencies, and
the extremely low detection limit that \emph{Kepler} has compared
to ground-based observations. And yet they have a resemblance
to the best-studied hybrid, BA09, in that these stars
all have intermediate frequencies, and BA09 also
has combination frequencies.

In these stars we detect 37, 41, and 21 frequencies for KIC002697388,
KIC003527751, and KPD~1943, respectively. In all cases, combination
frequencies may be present and the
pulsations were detected right down to
the detection limit, indicating that further data will reveal
more periodicities. All of these stars show widely distributed frequencies.
KIC002697388 and KIC003527751 show ``gap'' frequencies like BA09,
which lie between the $g-$ and $p-$mode regimes.

These stars will certainly be invaluable to discerning the interior
structure of sdB stars, as their pulsations span a large range of
frequencies, with each frequency, and frequency region likely probing
a different layer within the star itself. Such constraints can be
useful for determining if differential internal rotation exists as well
as probing narrow ionization (and thus convective) zones.

Typically, a key to discerning interior conditions of stars is the
association of frequencies with pulsation modes.
It is possible that the asymptotic approach, as has been applied to 
KIC010670103, or possible combination frequencies, as has been done for 
ZZ~Ceti stars \citep{clemens}, may be useful for
determining modes.
Such constraints could be very useful for distinguishing
between various models, modelling methods and help with mismatches
between observed and theoretical period distributions.

ACKNOWLEDGMENTS: Funding for this Discovery mission is provided by 
NASA's Science Mission
Directorate. The authors gratefully acknowledge the entire \emph{Kepler}
team, whose efforts have made these results possible. ACQ is supported by
the Missouri Space Grant Consortium, funded by NASA. 
The research leading to these results has received funding from the
European Research Council under the European Community's Seventh Framework
Programme (FP7/2007--2013)/ERC grant agreement n$^\circ$227224 (PROSPERITY) and
 from the Research Council of K.U.Leuven (GOA/2008/04). AB gratefully acknowledges
support from the Polish Ministry under grant No. 554/MOB/2009/0.

\end{document}